\documentstyle[11pt,paspconf,epsf]{article}
\begin{document}

\title{A Comparison of Simple Galaxy Mass Estimators}

\author{M. L. Weil\altaffilmark{1}}
\affil{Astronomy Department, Columbia University, New York, NY 10025}

\altaffiltext{1}{Department of Physics and Astronomy, McMaster University, 
Canada}

\section{Introduction}
Owing to their complex stellar dynamics, determining the mass profiles
and mass-to-light ratios of elliptical galaxies is difficult.  Now,
however, kinematic data, once limited to less than an effective
radius, are available out to several $R_{eff}$ and may include
higher-order moments of the velocity profile (e.g. Carollo et al. 1995,
Sembach \& Tonry 1996).  Because mean rotational velocities and
velocity dispersions reveal no information about the anisotropy of the
system, mass estimation methods which rely only on them are
degenerate.  Unfortunately, higher-order kinematical data are
difficult to obtain, the dynamical modeling applied to them remains
dependent on model assumptions, and the methods are time-consuming to
implement (e.g. Rix et al. 1997).

Here, three simple methods of mass estimation, which are easy to apply
to kinematic data, are compared.  
\begin{enumerate}
\item virial: $\displaystyle M(r) = {3 \pi \over 2G} \left(\sum_{i=1}^N 
\sigma_{i}^2\right)/\left(\sum_{i=1}^N {1 \over R_i} \right)$
\item projected: $\displaystyle M(r)={f_p \over \pi G N} \sum_{i=1}^N v_{rot,i}^2 R_i$, 
where $f_p=32$ for an isotropic orbital distribution ($f_p=64$ for radial
orbits).  
\item Jeans equation for a spherically symmetric system:
\end{enumerate}
\begin{equation}
{d(\rho v_r^2) \over dr} + {\rho \over r} (2 v_r^2 - 
v_{\vartheta}^2 - v_{\phi}^2) = -{GM(r) \rho \over r^2}
\end{equation}
in which second velocity moments $v_r^2=\sigma^2$,
$v_{\vartheta}^2=\sigma^2$, and $v_{\phi}^2=\sigma^2 + V_{rot}^2$
for an isotropic velocity dispersion.
Density, mass distribution, and rotation velocity are assumed: 
$\rho(r)={M_l a \over 2 \pi} {1 \over r(r+a)^3}$,
$M(r)={M_l r^2 \over (r+a)^2} + {M_d r^2 \over (r+d)^2}$,
$V_{rot}(r)={v_0 r \over (r^2 + r_0)^{1/2}}$,
where $M_l$ is luminous mass, $a$ is scale--length of the luminous matter,
$M_d$ is dark mass, $d$ is scale--length of the dark matter, and
$v_0$ and $r_0$ are determined by fitting the projected rotation to
the major axis of the remnants.

High-resolution models of elliptical galaxies
from N-body simulations are used to compare predicted mass profiles. 
Details of the eight models are found in Weil \& Hernquist (1994, 1996).  
The dark matter halos of the models are nearly spherical, with $<b/a>=0.97$ and
$<c/a>=0.86$.  The luminous components of the models are more triaxial,
with $<b/a>=0.88$ and $<c/a>=0.64$.

\section{Results}

The mass profile estimates for eight elliptical models were averaged
to compare the true and calculated mass profiles for the three mass
estimators.  For the Jeans equation mass estimator, projected rotation
$V_s(R)$, surface mass density $\Sigma(R)$, and the line-of-sight
second velocity moment $V_{los}^2(R)$ are determined such that
projected velocity dispersion is $\sigma_s^2(R)=V_{los}^2-V_s(R)$,
which depends on the four parameters which describe the mass
distribution, $a$, $M_l$, $d$, $M_d$ (code available from M.L. Weil).

Figure 1 shows the fractional difference profiles for
slits of length 34 kpc laid along projections of the stellar data. For
the virial and Jeans equation mass estimators, these are averaged over
six directions in the three intrinsic projections (short dashed and
solid lines).  For the projected and Jeans equation mass estimators,
results are also averaged over the two directions in which the true
rotational velocity, $v_{rot}$, appears (long dashed and long-short
dashed lines).  All the mass estimators are comparably poor for $r<4$
kpc, with errors greater than 30\%, and the projected estimator is
poor at all radii.  However, the errors for the other mass estimators
are within $\approx$ 10 - 20\% for $r>8$ kpc, although error for the Jeans
equation estimator, as applied to random projections, increases to 25\%
at $r=40$ kpc.  The Jeans equation estimator for projections with
$v_{rot}$ and the simple virial estimator reproduce the
true masses to within $\approx$ 15\% at large radii.

\begin{figure}
\plotone{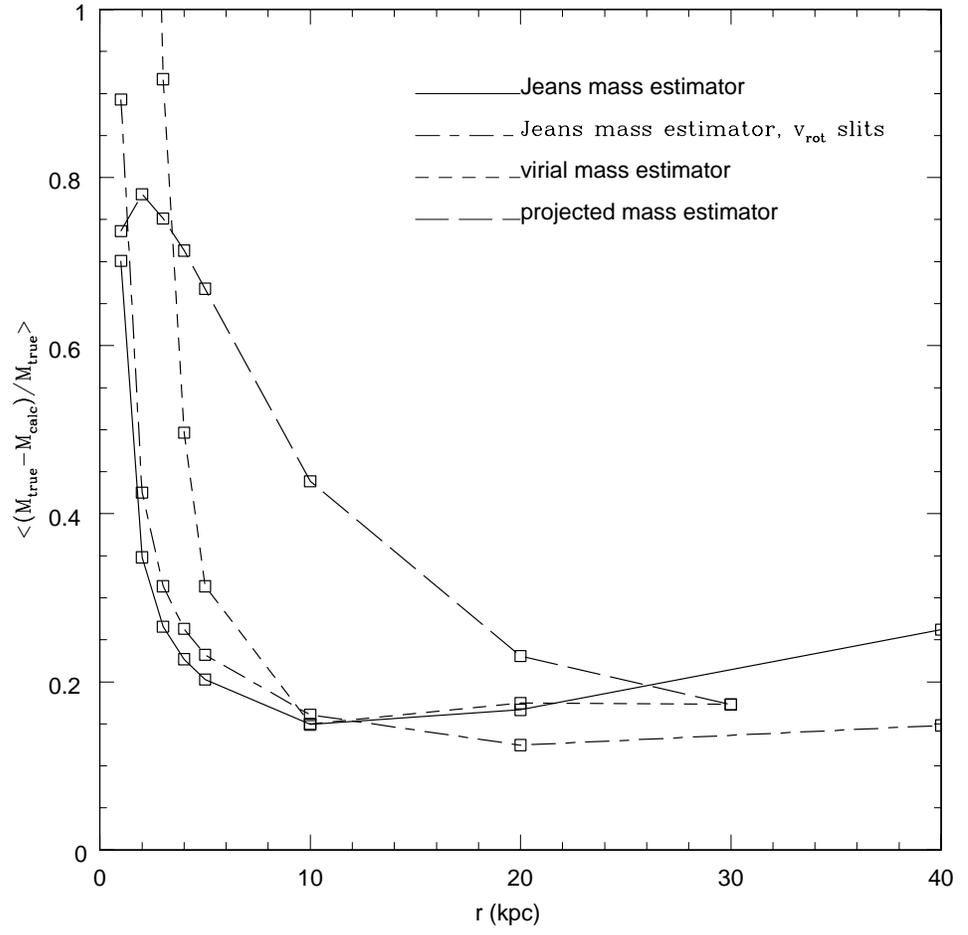}
\caption{Radial profile of fractional difference between true and calculated 
total masses for three mass estimators.} 
\label{figure1.ps}
\end{figure}

\end{document}